\documentclass[fleqn,twoside]{article}
\usepackage{espcrc2}

\usepackage{graphicx}

\usepackage[figuresright]{rotating}

\newcommand{\AmS}{{\protect\the\textfont2
  A\kern-.1667em\lower.5ex\hbox{M}\kern-.125emS}}

\title{The nuclear track detector CR39: results from different experiments}

\author{M. Giorgini\address[MCSD]{\small Dept. of Physics, University 
of Bologna, Viale C. Berti Pichat 6/2, I-40127 Bologna, Italy}
\address[MCSD]{\small INFN-Bologna, Viale C. Berti Pichat 6/2, I-40127 
Bologna, Italy \\ ~\\
{\normalsize Talk given at the $11^{th}$ Topical Seminar on Innovative 
Particle and Radiation Detectors,  Siena, Italy, 1-4 October 2008.}
}}

\begin{document}

\begin{abstract}
{\bf Abstract.} The nuclear track detector CR39 was
calibrated with different ions of different energies. Due to the low
detection threshold ($Z/\beta \sim 6e$) and the good charge resolution 
($\sigma_Z \sim 0.2e$ for $6e \leq Z/\beta \leq 83e$ with 2 measurements), 
the detector was used for different purposes: $(i)$ fragmentation of high 
and medium energy ions; $(ii)$ search for magnetic monopoles, nuclearites, 
 strangelets and Q-balls in the cosmic radiation.
\vspace{1pc}
\end{abstract}

\maketitle

\section{INTRODUCTION}
When an ionizing particle crosses a Nuclear Track Detector (NTD) sheet, it 
produces damages at the level of polymeric bonds around 
its trajectory, forming the so-called ``latent track''. For a 
particle with charge $Z$ and velocity $\beta=v/c$ the damage depends on 
the ratio $Z/\beta$. 

The chemical etching in a basic solution 
results in the formation of etch-pit cones in both faces of the sheet. The 
formation of etch-pit cones is regulated by the bulk etching rate, $v_B$, 
and the track etching rate, $v_T$, i.e. 
the velocities at which the undamaged and damaged materials are removed. 
Etch-pit cones are formed if $v_T$$>$$v_B$. The response of the 
detector is given by the ratio $p = v_T/v_B$ as a function of the 
Restricted Energy Loss (REL).

 The poly-allyl-diglycol carbonate polymer, commercially known as 
CR39, is the most sensitive NTD. 
It is sensitive for a wide range of charges down to $Z = 6e$ in the low 
velocity and in the relativistic regions \cite{framm,framm-2008,fading} and 
can reach a charge resolution $\sigma_Z \sim 0.05e$ \cite{hi-res}. CR39 was  
used to search for exotic particles, 
 like Magnetic Monopoles and Strange Quark Matter (SQM) 
\cite{slim-mm,macro-mm}, to 
study cosmic ray composition \cite{cake} and for environmental studies 
\cite{environ}. 

 In this paper the results of some applications of CR39 NTD are presented.

\begin{figure*}[htb]
\centering
{\includegraphics[height=5cm]{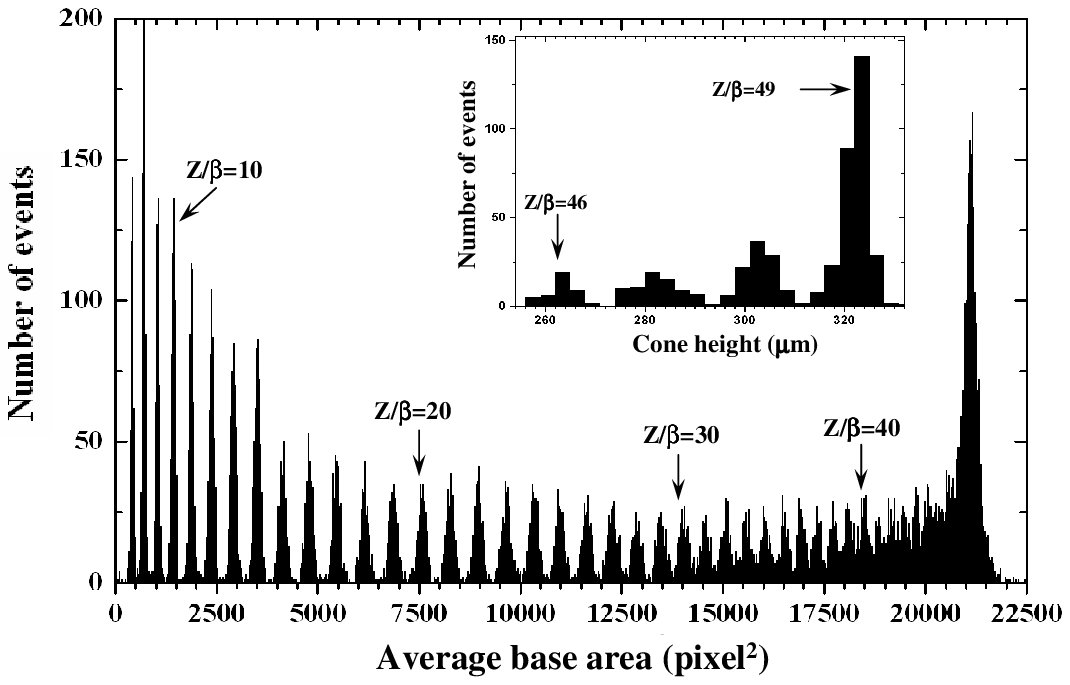}}
\hspace{3mm}
{\includegraphics[height=5cm]{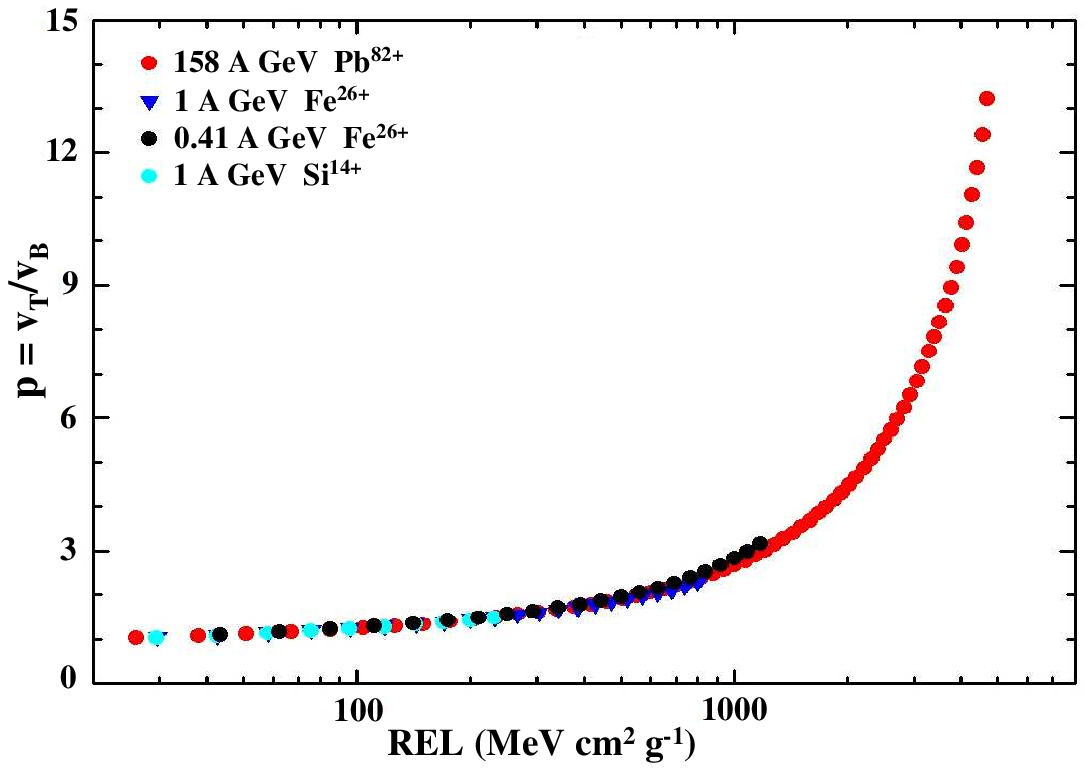}}
\vspace{-1cm}
\caption{(Left) Average (2 faces) base area distribution of etched cones 
in CR39 from In$^{49+}$ ions and their fragments 
($1~pixel^2 = 0.3~\mu$m$^2$). In the insert the cone 
height distribution for $46e \leq Z/\beta \leq 49e$ is shown. (Right) 
Summary of 
CR39 calibrations obtained with beams of different charge and energy.
Notice that a unique calibration curve describes all the data.}
\label{fig:calib}
 \end{figure*}

\section{CR39 CALIBRATIONS}

A stack composed of CR39 and Makrofol sheets of size $11.5 \times 11.5$ cm$^2$ 
with a 1 cm thick lead target was exposed to 158 A GeV Pb$^{82+}$ ions in 
1996; a second stack with a 1 cm thick aluminum target was exposed to 158 
A GeV In$^{49+}$ ions in 2003. Both exposures were performed at the 
CERN-SPS, at normal incidence and density of $\sim$2000 ions/cm$^2$. 
The detector sheets before and after the target recorded respectively 
the incident beam ions and the ones exiting from the target, together 
with their nuclear fragments.
 
After exposures, two CR39 sheets located after the target were etched 
in 6N NaOH + 1\% ethyl alcohol solution at 70 $^\circ$C for 40 h 
\cite{cal-CR39-Mak}. The addition of ethyl alcohol in the etchant improves the 
etched surface quality, reduces the number of surface defects and 
background tracks, increases the bulk etching velocity, speeds up the 
reaction, but raises the detection threshold.

For each CR39 sheet, the etch-pit base areas for beam ions 
and their fragments were measured with the Elbek automatic image analyzer 
system \cite{elbek}. Averages were computed from measurements made 
on the ``front sides'' of the detector sheets. The peaks are well separated 
from  $Z/\beta \sim 6e$  to $72e$ in the lead case \cite{ext-cal} and to $45e$ 
for the indium case \cite{cal-CR39-Mak}. The charge 
resolution close to the beam peak can be improved by measuring the heights of 
the etch-pit cones. The heights of 1000 etch-pit cones corresponding 
to nuclear 
fragments close to the beam peaks were measured with an accuracy of $\pm 1$ 
$\mu$m with a Leica microscope coupled to a CCD camera and a video monitor. 
The corresponding distributions \cite{cal-CR39-Mak,ext-cal} show 
that each peak is well separated from the others, and a charge can be 
assigned to each one. 
 Fig. \ref{fig:calib}(Left) shows the average (2 faces) base area 
distribution of etched cones in CR39 from In$^{49+}$ ions and 
their fragments. In the insert the cone 
height distribution for $46e \leq Z/\beta \leq 49e$ is shown.

For each detected nuclear fragment and for the
beam ions we computed the REL and the reduced etch rate $p = v_T/v_B$.
Fig. \ref{fig:calib}(Right) shows a summary of CR39 calibrations 
with different 
ions and energies. Notice that a unique calibration curve describes 
all the data.

\section{FRAGMENTATION CROSS SECTIONS OF {\boldmath {Fe$^{26+}$, Si$^{14+}$ 
AND C$^{6+}$}} IONS ON DIFFERENT TARGETS}

The availability of ion beams at the Brookhaven National 
Laboratory (BNL), USA, and at the Heavy Ion Medical Accelerator in 
Chiba (HIMAC), Japan, facilities made possible to investigate the projectile 
fragmentation on different targets and for different projectile energies.
 The present study is focused on Fe$^{26+}$, Si$^{14+}$ and C$^{6+}$ ion 
interactions in CH$_2$, CR39 and Al targets.

Stacks composed of several CR39 sheets of size $11.5 \times 11.5$ cm$^2$ and 
with different targets were exposed to 0.3, 1, 3, 5 and 10 A GeV 
Fe$^{26+}$, 1, 3, 5 A GeV Si$^{14+}$ ions at the BNL Alternating Gradient 
Synchrotron (AGS)
and NASA Space Radiation Laboratory (NSRL). The exposures to 0.41 A GeV 
Fe$^{26+}$ and 0.29 A GeV C$^{6+}$ ions were performed at HIMAC. 
For more details see \cite{framm-2008}.

\begin{figure*}[htb]
\centering
{\includegraphics[width=7.5cm,height=5cm]{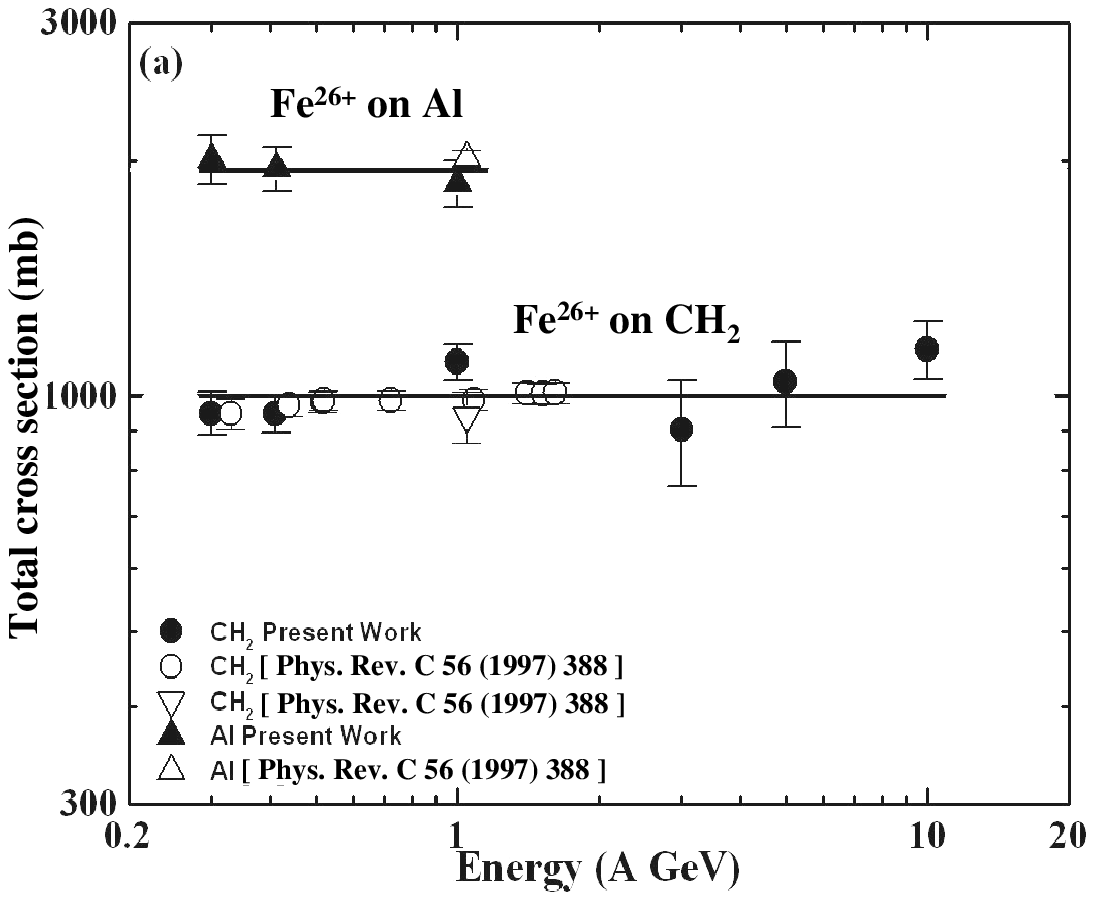}}
\hspace{0.5cm}
{\includegraphics[width=7.5cm,height=5cm]{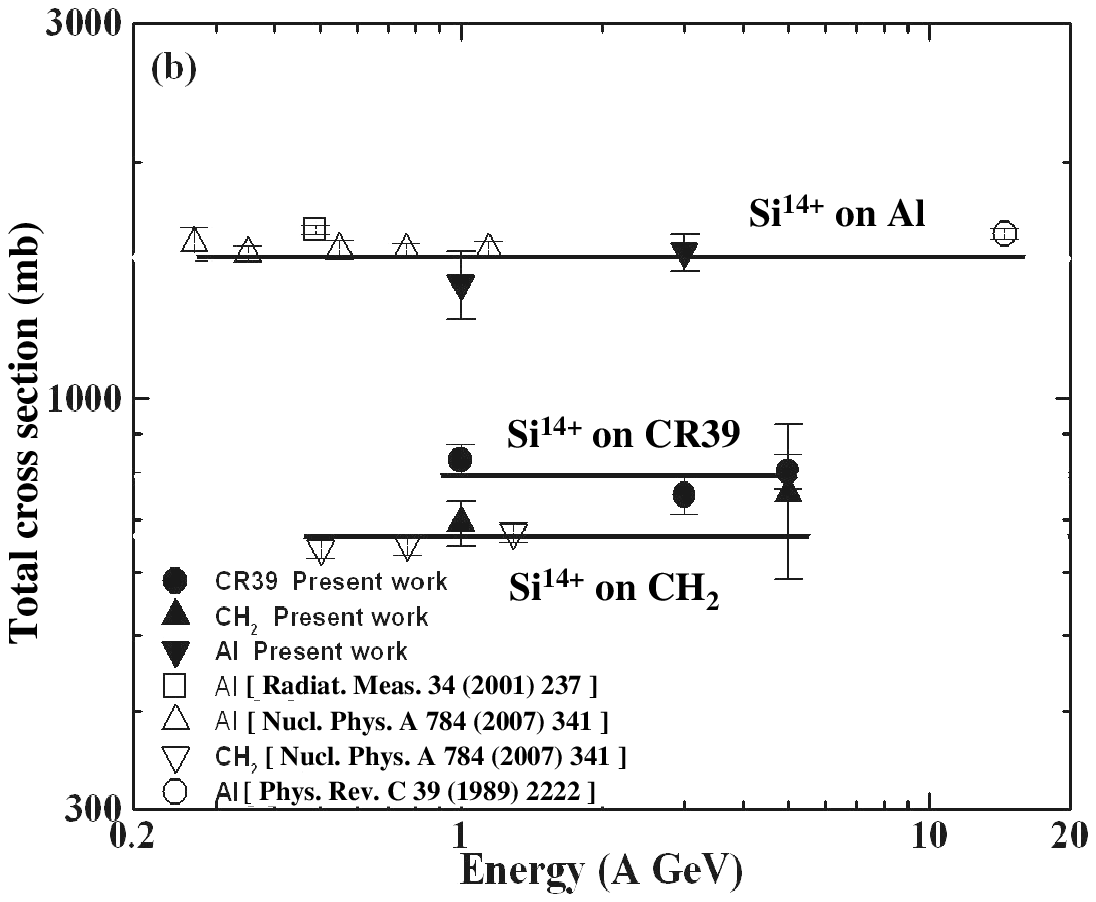}}
\vspace{-8mm}
\caption{Beam energy dependence of the total fragmentation cross sections of 
(a) Fe$^{26+}$ ions of 
  different energies in CH$_2$ and Al targets and (b) Si$^{14+}$ ions 
in CH$_2$, CR39 and Al targets. The measured cross sections from 
refs. \cite{framm-altri} (open symbols) and the predictions 
from the semi-empirical formula for nuclear cross sections 
\cite{bra-pet} (solid lines) are shown for 
comparison.} 
\label{fig:framm}
 \end{figure*}

We used three and four CR39 sheets, 
 $\sim$0.7 mm thick, placed before and after the target, respectively. The 
exposures were done at normal incidence, with a
density of $\sim$2000 ions/cm$^2$. After exposures, the CR39 sheets were 
etched in 6N NaOH aqueous solution at 70 $^\circ$C for 30 h.

The total charge changing cross sections were determined from the survival 
fraction of ions using the following relation
 \begin{equation}
 \sigma_{tot} = \frac {A_T \ln (N_{in} / N_{out})}{\rho~ t ~N_{Av}}
\end{equation}
where $A_T$ is the nuclear mass of the target (average nuclear mass in case 
of polymers: $A_{CH2} = 4.7,~ A_{CR39} = 7.4$); $N_{in}$ and $N_{out}$ are the 
numbers of incident ions before and after the target, respectively; $\rho$ 
(g/cm$^3$) is the target density; $t$ (cm) is the target thickness  and 
$N_{Av}$ is Avogadro number.

The average track base area was computed for each reconstructed ion path 
by requiring the existence of signals in at least two out of three detector  
sheets. The charge resolution $\sigma_Z$ was about $0.2e$. 
The numbers of incident and survived beam ions were determined considering 
the mean area distributions of the beam peaks before and after the target 
and evaluating the integral of the gaussian fit of the beam peaks.

 Fig. \ref{fig:framm} shows the beam energy dependence of the total 
charge changing cross sections of (a) Fe$^{26+}$ projectiles on the CH$_2$ 
and Al targets and (b) Si$^{14+}$  projectiles on the CH$_2$, CR39 
and Al targets. Our data are almost energy independent, in agreement 
with the data from refs. \cite{framm-altri} and with the predictions of 
the semi-empirical formula \cite{bra-pet} for nuclear cross sections 
(solid lines).

\section{SEARCH FOR RARE PARTICLES}
The SLIM (Search for LIght Monopoles) experiment was an array of NTDs 
with a total area of 427 m$^2$ exposed for 4.22 years 
 at the Chacaltaya laboratory in Bolivia at 5230 m a.s.l.
\cite{slim-mm,proposal}. The array 
was organized into 7410 modules, each of area 
$24 \times 24$ cm$^2$. All modules were made up of: three layers of 
CR39, each 1.4 mm thick; 3 layers of Makrofol, each 0.48 mm thick; 2 
layers of Lexan, each 0.25 mm thick, and one layer of aluminum absorber 
1 mm thick. 

The main purpose of the SLIM experiment
 was the search for Intermediate Mass Monopoles (IMMs) with masses 
$10^{5}< M_{M} < 10^{12}$ GeV \cite{IMMs}. These IMMs, predicted by 
some GUT and supersymmetric models, may have been
produced in the early Universe and could be 
present in the cosmic radiation \cite{UHECR}. 
The SLIM detector was also sensitive to Strange Quark Matter 
nuggets \cite{nuclr} and Q-balls \cite{qballs}.

 Since IMMs have a constant energy loss through the 
stacks, the subsequent chemical etching should result in collinear 
etch-pit cones of equal size on both faces of each detector sheet. 
 In order to increase the  detector ``signal to noise'' ratio different 
etching conditions \cite{fading,cal-CR39-Mak,ext-cal} 
were defined.
 The {\it strong etching} (8N KOH + 1.5\% ethyl alcohol at 
75 $^\circ$C for 30 hours) allows better surface quality 
and larger post-etched cones to be obtained. This makes etch-pits easier 
to detect under visual scanning. 
The {\it soft etching} (6N NaOH + 1\% ethyl alcohol at 70 $^\circ$C 
for 40 hours) allows to 
proceed in several etching steps and study the formation of the 
post-etched cones.

\begin{figure}[htb]
\centering
{\includegraphics[height=4.5cm]{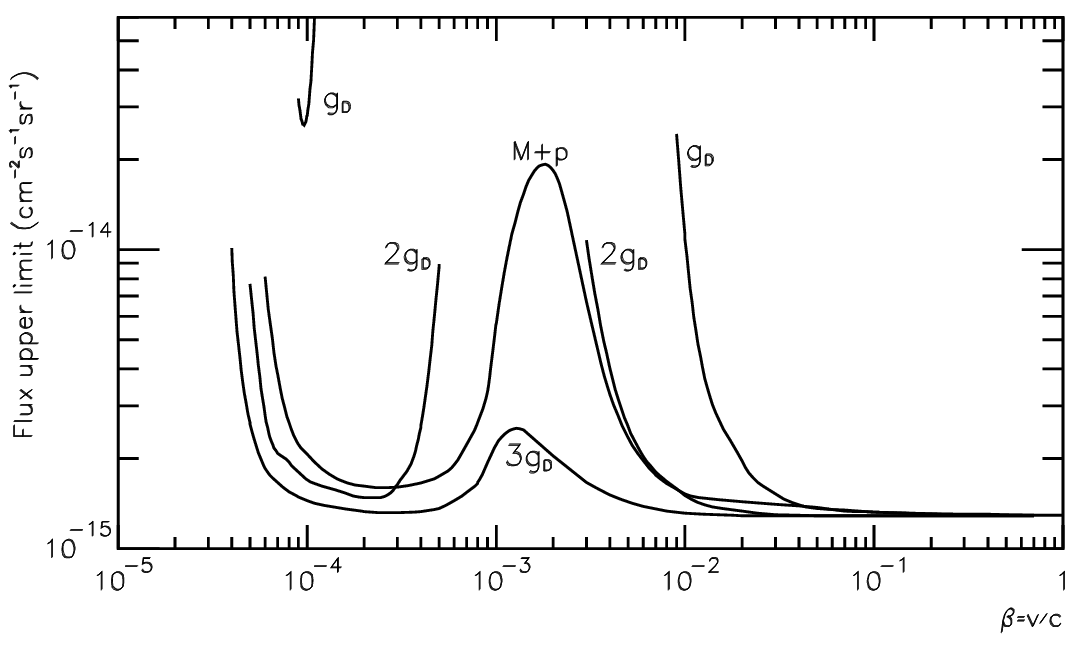}}
\vspace{-1.5cm}
\caption{90\% C.L. upper limits for a downgoing
 flux of IMMs with $g=g_D,~2g_D,~3g_D$ and for dyons (M+p, $g=g_D$) 
plotted vs $\beta$.}
\label{fig:limite}
 \end{figure}

The analysis of a SLIM module started by etching the uppermost CR39 sheet 
using strong conditions in order to reduce the CR39 thickness from 
1.4 mm to $\sim 0.9$ mm. After the strong etching, the CR39 sheet was scanned 
twice, with a stereo microscope, by different operators, 
 with a 3$\times$ magnification optical lens, looking for any possible 
correspondence of etch pits on the two opposite surfaces. The measured 
single scan efficiency was about 99\%; thus the double scan guarantees an 
efficiency of $\sim 100\%$ for finding a possible signal.  

 A track was defined as  a ``candidate'' if the computed momentum  
and incident angle on the front and back sides were equal to within 20\%.
 In this case the lowermost CR39 layer was etched in soft etching 
conditions, and an accurate scan under an optical microscope with high 
magnification (500$\times$ or 1000$\times$) was 
performed in a square region around the candidate expected position, 
which included  the ``coincidence'' area ($\sim$0.5 cm$^2$). 
If a two-fold coincidence was detected, the CR39 middle layer was also 
analyzed. 

Since no candidates were found, the global 90\% C.L. upper limits for the 
flux of downgoing IMMs and dyons with velocities $\beta > 4 \cdot 10^{-5}$ 
were computed, as shown in Fig. 
\ref{fig:limite}. The flux limit for $\beta > 3 \cdot 10^{-2}$ is 
$\sim 1.3 \cdot 10^{-15}$ cm$^{-2}$ s$^{-1}$ sr$^{-1}$ \cite{slim-mm}.    

The 90\% C.L. upper limits for a downgoing flux of nuclearites and charged 
Q-balls is $\sim 1.3 \cdot 10^{-15}$ cm$^{-2}$ s$^{-1}$ sr$^{-1}$ for   
$\beta >  10^{-4}$ \cite{slim-sqm}.

\section{CONCLUSIONS} 

The nuclear track detector CR39 was calibrated with different ions of different
energies. A unique curve of $p$ vs REL describes all the data.

The total fragmentation cross sections for Fe$^{26+}$, 
Si$^{14+}$ and C$^{6+}$ ions on
polyethylene, CR39 and aluminum targets were measured using CR39.
The total cross sections 
do not show any observable energy dependence and are in agreement 
with similar data in the literature.

In the SLIM experiment we etched and analyzed 427 m$^2$ of CR39. No 
candidate passed the search criteria. The 90\%
C.L. upper limits for a downgoing flux of fast ($\beta > 3 \cdot 10^{-2}$) 
downgoing IMMs are at the level of $1.3 \cdot 10^{-15}$ cm$^{-2}$ 
sr$^{-1}$ s$^{-1}$. The same limits were obtained for nuclearites, strangelets 
and charged Q-balls with $\beta > 10^{-4}$.

\end{document}